\documentclass[aps,prl,twocolumn,showpacs,superscriptaddress,amsmath,amssymb]{revtex4}

\usepackage{graphicx}
\usepackage{dcolumn}
\usepackage{bm}

\begin{document}

\title{Understanding the Nucleation Mechanisms of Carbon Nanotubes in
  Catalytic Chemical Vapor Deposition.}

\author{H. Amara} \affiliation{Laboratoire de Physique du Solide, Facult\'es
  Universitaires Notre-Dame de la Paix, 61 Rue de Bruxelles, 5000
  Namur, Belgique} \affiliation{PCPM, Universit\'e Catholique de
  Louvain, Place Croix du Sud 1, 1348 Louvain-la Neuve, Belgique}

\author{C. Bichara}
\affiliation{Centre de Recherches en Matiere Condens\'ee et Nanosciences, CRMCN-CNRS, Campus de Luminy,
Case 913, 13288 Marseille Cedex 09, France.}

\author{F. Ducastelle}
\affiliation{Laboratoire d'Etudes des Microstructures,
ONERA-CNRS, BP 72, 92322
Ch\^atillon Cedex, France.}

\date{\today}

\begin{abstract}

  The nucleation of carbon caps on small nickel clusters is studied using a tight binding model coupled to grand canonical Monte Carlo simulations. It takes place in a well defined carbon chemical potential range, when a critical concentration of surface carbon atoms is reached. The solubility of carbon in the outermost Ni layers, that depends on the initial, crystalline or disordered, state of the catalyst and on the thermodynamic conditions, is therefore a key quantity to control the nucleation.
\end{abstract}


\maketitle

Although multiwall carbon nanotubes are now produced on an industrial scale  \cite{Grobert_2007}, a detailed understanding of the synthesis mechanisms of single wall carbon nanotubes (SWNT) is still mandatory for more elaborate applications \cite{Dresselhaus2001}. Diameter analysis of the tubes produced by Chemical Vapor Deposition (CVD) \cite{Nasibulin_2005} and a tendency towards chiral selectivity \cite{Lolli, Bachilo_2003} have been reported, but simple questions such as the role and physical state of the catalyst are still not answered. Transmission Electron Microscopy (TEM) observations performed after the synthesis lead to contradictory results showing evidences for either crystalline \cite{Zhu_small} or liquid state \cite{Harutyunyan_2005} of the metal particle during  the synthesis. Remarkable progress has been made in the \emph{in situ} observation of the growth \cite{Helveg,  Lin_Nanoletters_2006, Hofmann_NL_2007, Rodriguez_2007}, but the atomic resolution is not yet obtained under these conditions.\\
Computer simulation is then a unique tool to gain an insight at this atomic scale that is very difficult to access experimentally. However, obtaining an accurate description of the interatomic interactions, as typically provided by first principles calculations, and addressing the size and time scales relevant to the experiments are extremely difficult challenges. Efficient catalyst particles for CVD are in the 1-10 nm diameter range and growth rates are in the nano- to micrometer per second range \cite{Dai_2002, Lin_Nanoletters_2006}. Such scales are largely beyond the capabilities of Molecular Dynamics (MD) techniques, meaning that only some elementary steps can be studied, or that the growth conditions imposed in the simulations are orders of magnitude too fast.\\
Static \textit{ab initio} calculations were used to study the carbon/catalyst interaction \cite{Fan2003, Zhang2004, Abild2006}. Hofmann \emph{et al.} \cite{Hofmann2005} and Abild-Pedersen \emph{et al.} \cite{Abild2006} studied surface or subsurface C diffusion on Ni, and schematic growth model for CVD were proposed \cite{Puretzky_2005, Hofmann2005}. First principles MD simulations were used to study the root growth on a Co catalyst \cite{Gavillet2001} and the nucleation of a C cap on a Fe cluster was described by Raty \emph{et al.} \cite{Raty2005}. The latter explained the absence of diffusion inside the Fe particle by its nano size (55 atoms). However Ding \emph{et al.} \cite{Ding2006}, using a semi empirical FeC interaction model, found that nucleating C islands on a Fe particle required a large bulk C supersaturation. Shibuta \emph{et al.} \cite{Shibuta2003} also found a large C solubility in small Ni clusters, in contradiction with the \emph{ab initio} calculations of Zhang \emph{et al.} \cite{Zhang2004} showing that, at 0 K, C segregation at the surface of a Ni$_{38}$ cluster is preferred. Studying the growth of nanotubes on small (Ni$_{48}$ and Ni$_{80}$) clusters by empirical MD, Zhao \emph{et al.} \cite{Balbuena_Nanotechnology} found the same mechanism as in our previous calculations on a semi infinite system (slab geometry) \cite{Hakim2006} based on a tight binding model. These differences are partly due to the interatomic interaction model used but we show below that the initial state of the catalyst is an important parameter. Moreover, the MD technique itself is not able to take into account the carbon chemical potential ($\mu_{C}$) gradient that is the thermodynamic driving force for the growth.\\
In this Letter, we adopt a new approach, using a tight binding model for Ni-C interactions coupled to grand canonical Monte Carlo (GCMC) simulations, to gain a new insight on the of nucleation of C nanotubes. Experimentally, $\mu_{C}$ is imposed by the thermochemistry of the catalytic decomposition reaction of the carbon rich feedstock, which is the first step of all CVD processes. Introducing $\mu_{C}$ as a control variable of the computer simulations requires using a GCMC algorithm, with fixed volume, temperature, number of Ni atoms and $\mu_{C}$ \cite{Frenkel1996}. Such simulations are long and proceed by sequentially attempting random changes of the atomic configurations: they have to rely on a model in which the total energy is taken as a sum of local terms, to avoid recalculating it for the complete simulation box at each step.\\
We developed and carefully tested \cite{Amara2005, Hakim2006} such a model for Ni and C, in a tight binding framework. The total energy is a sum of local terms: an empirical repulsive one and a band structure one including \textit{s} and \textit{p} electrons of C and \textit{d} electrons of Ni. Local densities of electronic states (LDOS) are calculated using the recursion method. To keep the model as simple and fast to compute as possible, we neglect the Ni \textit{s} electrons and calculate only the first four moments of the LDOS. The energy of each atom therefore depends only on the positions and chemical identities of its first and second neighbors, as defined by a cut off distance set at 3.20 \AA\ . \\
In the GCMC algorithm used here, all atoms are allowed to move while only C atoms are tentatively added or deleted in a volume close (3 \AA\ above or below) to the surface of the Ni particle. Moreover, the lowest part (typically 40 $\%$) of the particle is excluded for attempted insertions, in order to simulate a supported cluster and to avoid a possible encapsulation of the particle. More details on this algorithm that closely follows the thermodynamic conditions of CVD synthesis can be found in \cite{Hakim_JNN2007}. To analyze the structures, we distinguish three types of C atoms, according to the type (C or Ni) and length ($d_{CC}$ or $d_{CNi}$) of their nearest neighbor bonds. Outer C atoms, that ultimately form a nanotube cap, are defined as C atoms with at least one C neighbor at $d_{CC}<1.70$ \AA\ . The remaining C atoms are subdivided into surface or bulk atoms, according to their number of Ni neighbors, defined by $d_{CNi}<2.30$ \AA\ : those with less than 5 Ni neighbors are defined as surface C atoms, those with more than 5 neighbors as bulk C atoms. Note that these so called bulk atoms are essentially located in subsurface sites, but a diffusion towards interstitial sites buried deeper in the system is sometimes observed.\\
\begin{figure}[htbp!]
\includegraphics[width=8cm]{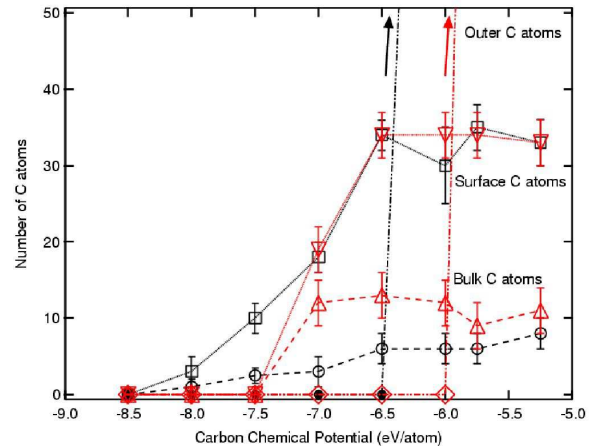}
\caption{Adsorption isotherms calculated on crystalline (black) or disordered (red) Ni$_{201}$ atoms clusters. Open circles and up pointing triangles correspond to C atoms dissolved in the bulk of the cluster; open squares and down pointing triangles to surface C atoms; full circles and open diamonds to outer C atoms. See text for the definition of these three kind of species. The arrows indicate that, beyond $\mu_{C}$ = -6.50 or -6.00 eV/at., the number of outer C atoms keeps increasing, while the other C species tend to equilibrium values.}
\label{Figure_1}
\end{figure}
The adsorption isotherms calculated at 1000 K, a typical temperature for CVD processes, are displayed in Fig. \ref{Figure_1}. Two clusters of 201 Ni atoms were considered : a crystalline one, in its equilibrium Wulff shape, and a disordered one, resulting from the quench of a liquid droplet. The $\mu_{C}$ values of interest for C adsorption are close to the energy of formation of graphene or diamond (-7.41 and -7.39 eV/atom in our model) and above. Adsorption begins at a lower $\mu_{C}$ value (-8.00 eV/atom) on the FCC cluster than on the disordered one (-7.50 eV/atom) because the semi octahedral sites of the (100) facets of the former are very favorable. Above these values, and below a critical value $\mu_{C}^{*}$ that depends on the state of the Ni cluster, an increasingly large number of C atoms is adsorbed on the cluster surface or in subsurface interstitial sites. This $\mu_{C}^{*}$ value at which C atoms begin to form structures outside the cluster lies between -6.50 and -6.00 eV/atom for the crystalline cluster and between -6.00 and -5.75 eV/atom for the disordered one: below $\mu_{C}^{*}$ the number of outer C atoms is zero, while a very steep increase is noticed at the transition. Around $\mu_{C}^{*}$ and beyond, the number of bulk and surface C atoms reaches a constant equilibrium value. We note that the number of bulk C atoms is larger in the disordered Ni cluster (11$\pm$3) than in the crystalline one (6$\pm$2). This difference, observed on a nanosized object, is qualitatively the same as what is known for bulk phase diagrams: at constant temperature, a liquid phase generally exhibits a larger solubility than the corresponding crystalline one. On the contrary, within the statistical error bars, the number of surface C atoms (34$\pm$3) does not depend on the state of the Ni particle. Referred to the 81 surface Ni atoms of the active part of the crystalline cluster, this corresponds to a critical surface C concentration of about 30 $\%$. Canonical Monte Carlo calculations at 1000 K, starting with a fixed number of C atoms randomly deposited on a Ni(111) surface, show that this critical surface concentration is large enough to allow some C atoms to interact with C neighbors and form small chains (with 3 to 5 C atoms) that are stable over relatively long 'time' scales. Under grand canonical conditions, these chains act as nucleation sites onto which incoming C atoms attach. When the catalyst particle is disordered, the diffusion of C atoms towards subsurface sites is easier and the stability of these small chain embryos is lower, explaining the slightly larger $\mu_{C}^{*}$ value for disordered clusters. \\
\begin{figure*}[htbp!]
\begin{center}
\includegraphics[width=0.99\linewidth]{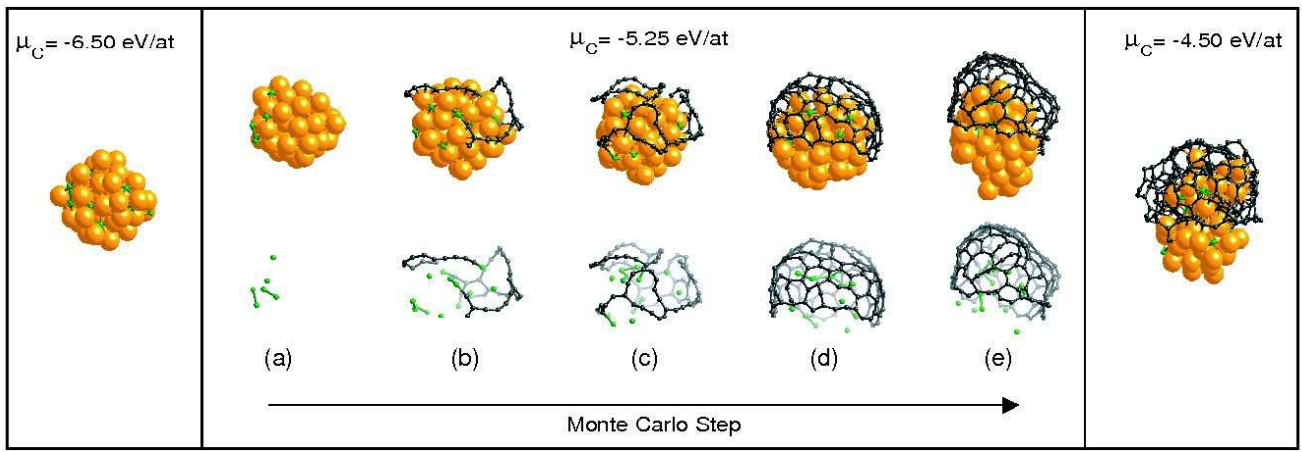}
\end{center}
\caption{GCMC calculations on a Ni$_{50}$ cluster at 1000 K. Ni atoms are orange, outer C atoms are black, surface or bulk C atoms are green. Left panel: Final configuration obtained for $\mu_{C}$ = -6.50 eV : no growth is observed. Central panel : different stages of the nucleation of a C cap at $\mu_{C}$ = -5.25 eV/at.. (a) atoms or dimers adsorbed; (b) chains forming; (c) chains crossing; (d) cap formed; (e) cap lifts off. Right panel : final configuration obtained for $\mu_{C}$ = -4.50 eV resulting in the formation of an amorphous C layer.}
\label{Figure_2}
\end{figure*}
We now focus on the outer C atoms. Figure \ref{Figure_2} presents a series of snapshots taken at different stages of the nucleation of a nanotube cap on a small disordered cluster of 50 Ni atoms, obtained by GCMC calculations at 1000 K and $\mu_{C}$ = -5.25 eV/atom. When the critical number of surface C atoms is reached, chains are formed, crawl on the surface and eventually cross each other. At their intersections, three fold coordinated C atoms act as nucleation centers for C $sp^{2}$ structures that develop on the surface. These $sp^{2}$ C atoms interact weakly with the underlying Ni atoms and can detach from the surface. The energy of adhesion of a perfect graphene layer in epitaxy on a Ni(111) surface has indeed been shown to be close to zero \cite{Abild2006,Hakim2006}. This growth pattern was already observed on flat Ni slabs \cite{Hakim2006}. The difference is that the curvature necessary to form a nanotube cap is now provided by the curvature of the small Ni cluster, meaning that the diameter of the nanotube cap corresponds to the local curvature of the Ni particle, at the moment of the nucleation, as observed in the \emph{in situ} TEM observations \cite{Hofmann_NL_2007}. In addition, the systematic analysis of the role of $\mu_{C}$ shows that there is an optimal $\mu_{C}$ range for nucleating nanotubes. The existence of the lower threshold ($\mu_{C}^{*}$) is in qualitative agreement with the conclusions of Lolli \emph{et al.} \cite{Lolli}, who showed that a lowering of the C surface fugacity hinders the nucleation of C species on the metal surface. Beyond the upper threshold (roughly, $\mu_{C} \geq -5.00$ eV/atom), the fast growth of an outer C structure leads to the formation of amorphous carbon (ultimately carbon fibers), similar to those observed by Helveg \emph{et al.} \cite{Helveg}.\\
\begin{figure}[htbp!]
\includegraphics[width=8cm]{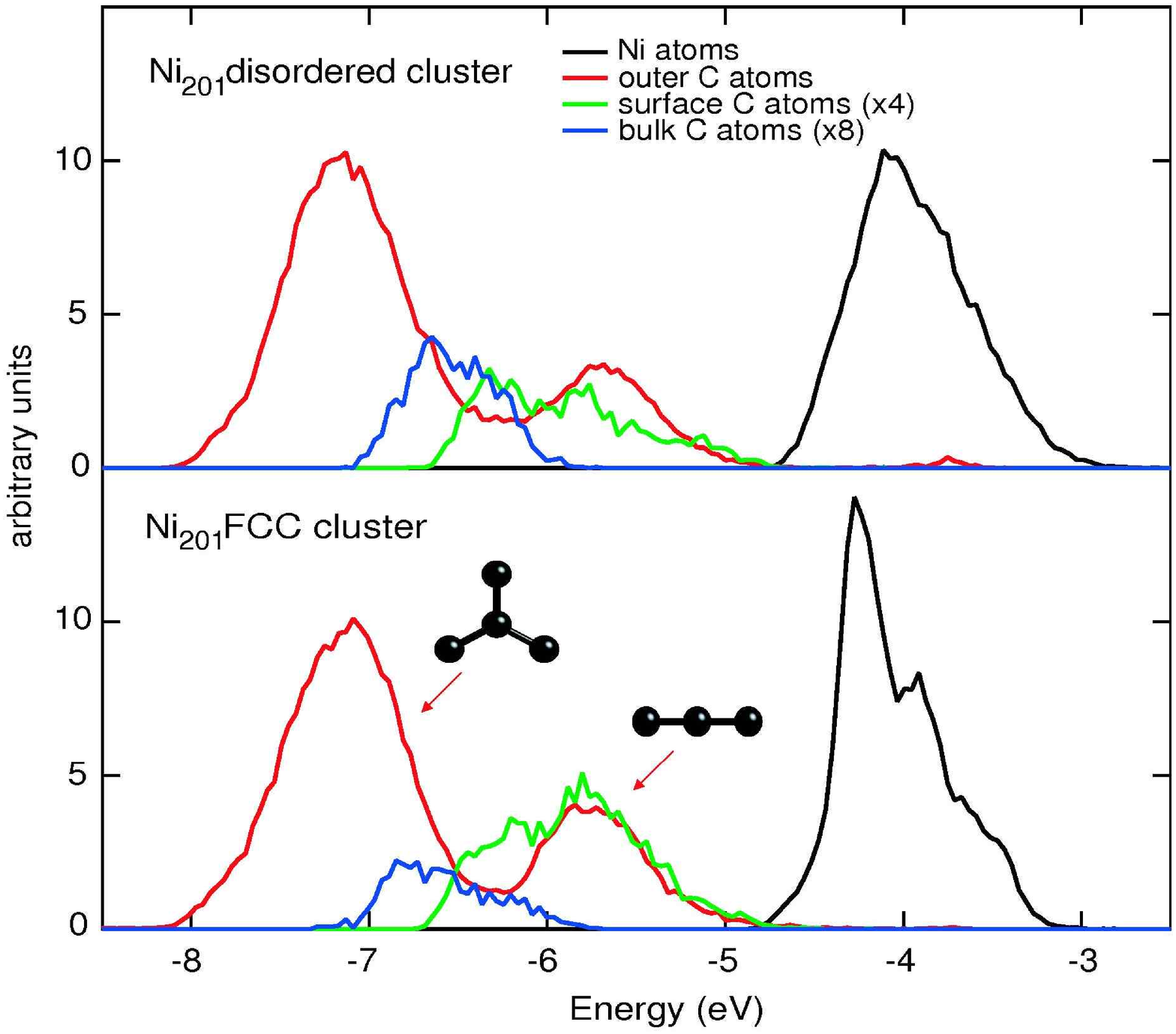}
\caption{Energy distribution of each species, calculated on Ni$_{201}$ crystalline (lower panel) or disordered (upper panel). Note the bimodal shape of the outer C atoms distribution.}
\label{Figure_3}
\end{figure}
We now understand the role of the catalyst, beyond its primary, truly catalytic, activity in the gas feed decomposition reaction. The above analysis of the partition of adsorbed C atoms into bulk, surface or outer sites shows that the first role of the metal catalyst is to confine atomic C on, or close to the surface until a critical concentration is reached. The local energies distributions plotted in Fig. \ref{Figure_3} tell more. The energy distribution of Ni atoms is centered around -4.0 eV and that of outer C atoms is bimodal: the peak around -5.8 eV corresponds to C atoms with 2 C neighbors, while the peak around -7.2 eV corresponds to the more stable threefold coordinated C atoms. Using \emph{in situ} XPS spectroscopy Hofmann \emph{et al.} \cite{Hofmann_NL_2007} evidenced the presence of chemisorbed C close to the surface of the Fe catalyst, stable during $\sim$ 150 s, prior to the formation of graphitic C. We also observe this and note that the energy of individual C atoms adsorbed on the surface matches that of the twofold coordinated outer C atoms: this is the second role of the metal catalyst. As illustrated in Fig. \ref{Figure_2}, it makes the transition from isolated atoms or dimers adsorbed on the surface to chains of $sp$ C possible with a low energy cost, as soon as the critical concentration is reached. This transition is a continuous one, without activation energy needed. As also shown in \cite{Amara2005, Abild2006} subsurface C atoms are only slightly more stable than surface ones, otherwise the tendency to form a carbide would prevent the surface segregation of C. Deck and Vecchio \cite{Deck_2006} have shown experimentally that successful catalysts have C solubility limits between 0.5 wt.$\%$ and 1.5 wt.$\%$ C. The energy pattern observed is compatible with such a limited solubility range and is the key for nucleating C nanotubes.\\
\begin{figure}[htbp!]
\includegraphics[width=8cm]{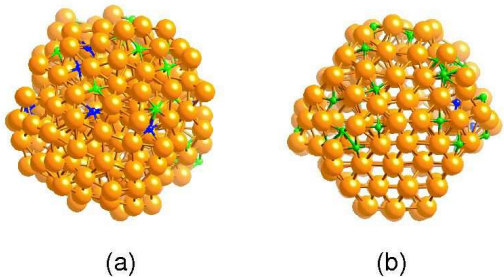}
\caption{Final state of disordered (a) or initially crystalline (b) Ni$_{201}$ catalyst particles, after formation of an outer C layer that has been removed for legibility. Note that the latter remains essentially crystalline, except for the outermost layer. Green atoms are surface C atoms, blues ones are bulk C atoms. See text for definition of these two kinds of species.}
\label{Figure_4}
\end{figure}
As shown in Fig. \ref{Figure_4}, the bulk or subsurface carbon concentration depends on the state (crystalline or disordered) of the catalyst. The initially crystalline cluster retains a large degree of crystallinity, but the observed preference for incorporating C atoms in interstitial sites located below edge or surface Ni atoms induces a strong local disorder. As shown in \cite{Hakim2006}, a large concentration of C in the outermost Ni layers induces a surface melting. Depending on the initial state of the particle, the temperature and the chemical potential, a molten surface layer may exist, making a fast diffusion of both Ni and C possible and possibly explaining the reshaping of the catalyst particle observed by \emph{in situ} TEM experiments \cite{Helveg, Hofmann_NL_2007}.  We thus reconcile conflicting experimental observations : some TEM observations (Zhu \emph{et al.} \cite{Zhu_small}, for example) show clearly crystalline particles, while others (e.g.: Harutyunyan \emph{et al.} \cite{Harut_2005}) and time resolved reflectivity measurements by Puretzky et al. \cite{Puretzky_2005} support the presence of a molten or partially molten catalytic particle.\\
In this Letter we have shown that an optimal $\mu_{C}$ window exists to nucleate nanotube caps whose curvature match the local curvature of the catalyst particle. The nucleation is triggered when the concentration of C atoms adsorbed on the surface is large enough, about 30 $\%$ in our calculations. The role of the catalyst is to confine C atoms on or close to the surface, and to make them reach this critical concentration. Moreover, the energies of isolated C atoms on the surface and of C chains adsorbed are in the same range, making a continuous transition possible. The C concentration in bulk and subsurface layers depends on the state (crystalline or disordered) of the catalyst.
Beyond their agreement with most experimental observations, these results emphasize the importance of the limited C solubility in Ni and give an insight at the atomic level on the structure of the tube/catalyst interface. They represent an essential step towards a better control of the structure (diameter, chirality) of the tubes formed and should help develop better catalysts.\\

\end{document}